# Advanced Mathematical Business Strategy Formulation Design

Song-Kyoo (Amang) Kim

**Abstract**

This paper deals with the explicit design of strategy formulations to make the best strategic choices from a conventional matrix form of representing strategic choices. The explicit strategy formulation is an analytical model which is targeted to provide a mathematical strategy framework to find the best moment for strategy shifting to prepare rapid market changes. This theoretical model could be adapted into practically any strategic decision making situation when a strategic formulation is described as a matrix form with quantitative measured decision parameters. Analytically tractable results are obtained by using the fluctuation theory and these results are capable to predict the best moments of changing strategies in a matrix form. This research helps strategy decision makers who want to find the optimal moments of shifting present strategies.

## 1. Introduction

The strategy is a set of actions which contains the execution of core competencies and the collection of competitive advantages. The determination of long-run goals, the adaptation for series of actions, and the allocation of resources are handled for carrying out strategic goals. Hence, companies could make choices among competing alternatives as the pathway to pursue strategic competitiveness by choosing proper strategies. The strategic management is operating the full set of commitments, decisions, and actions for gaining break-even returns by achieving strategic competitiveness [1-3]. The strategy formulation is a particular process to choose the most appropriate strategic actions for realizing objectives and for achieving visions of a company [4]. Formulation results provide a blueprint of strategic actions to achieve goals of a company. A strategy formulation forces a company to find the moments of environment changes and to be ready for shifting strategies [5]. A conventional strategic formulation includes defining a corporate mission, specifying achievable objectives, developing strategies, and setting policy guidelines [3]. It is also based on the sources of decision parameters which could develop visions and missions of a company by formulating one or more strategies with available information [1]. The main contribution provides the general framework for formulating the strategy which is explicitly designed and mathematically proven and the theory to determine the explicit probability of the strategy shifting is also included. This mathematical model explicitly identifies the best moment of shifting strategic choices and the analytical form could even count one step prior to pass the thresholds of strategic decision parameters. The fluctuation theory is one of powerful mathematical tools in the stochastic modeling and the first exceed model is a variation of conventional fluctuation models [12, 13]. In the fluctuation model [12, 13], the multi-compound renewal process evolves until either of the components hit (i.e. reach or exceed) their assigned levels for the first moment and the associated random variables which include the first passage time, the first passage level and the termination index. The first exceed level theory in the fluctuation model has been applied into various applications including the antagonistic games [14-16], the stochastic defense system [17], the Blockchain Governance Games [18, 19] and the versatile stochastic duel games [20, 21]. Analytically solved probability distributions by using the variation of the fluctuation model [13-15] is one of core contributions of this paper.

## 2. LITERATURE REVIEW

A strategy formulation helps a company to prepare the moment of shifting strategies. As mentioned in the previous section, strategic formulation tools include the business level strategy by Porter [7], the product-market matrix by Ansoff [9] and the growth-share matrix by the Boston Consulting Group [22,23]. The purpose of this section is showing how simple matrix form could be adapted into various strategic formulations intuitively. It is noted that actual applications for using strategic formulations will not be covered in this section.

**Strategic Group Mapping:** Strategic groups are groups of which team members are gathered by similar strategic characteristics, following similar strategies within an industry or a sector [2]. The strategic group matrix maps based on the R&D intensity and export focus that distinguish between strategic groups (see Fig. 1).

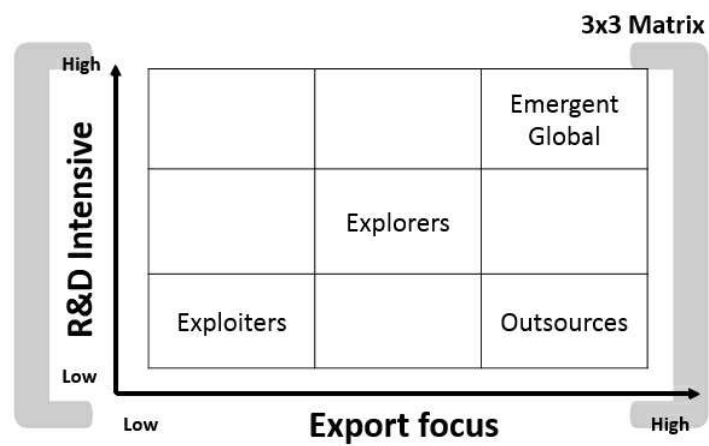

**Fig. 1.** Strategic Group Mapping [2]

**Issues Priority Matrix** is the matrix form to identify and to analyze developments in the external environment [3, 6]. The issues priority matrix is applied for helping managers to decide which environmental trends should be merely scanned (low priority) and which should be monitored as strategic factors (high priority). The decision parameters are the probable impacts and the chances of occurrences (see Fig. 4).

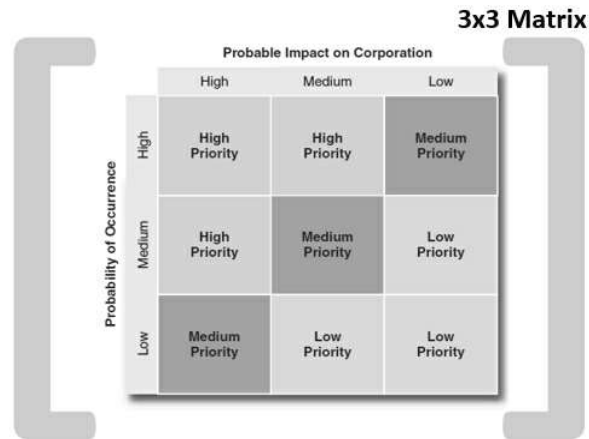

**Fig. 4.** Issues Priority Matrix adapted from Lederman [7]

**Value-Creation Diversification Matrix** is a matrix form of a corporate-level strategy specifies actions [1]. Corporate-level strategies help firms to select new strategic positions and the matrix specifies actions of company to gain a competitive advantage by selecting a proper strategy. Operational relatedness and corporate relatedness are the decision parameters to determine a best strategy. The decision parameter of the vertical dimension is an opportunity to share operational activities between businesses while the horizontal dimension suggests an opportunity for transferring corporate-level core competencies (see Fig. 5).

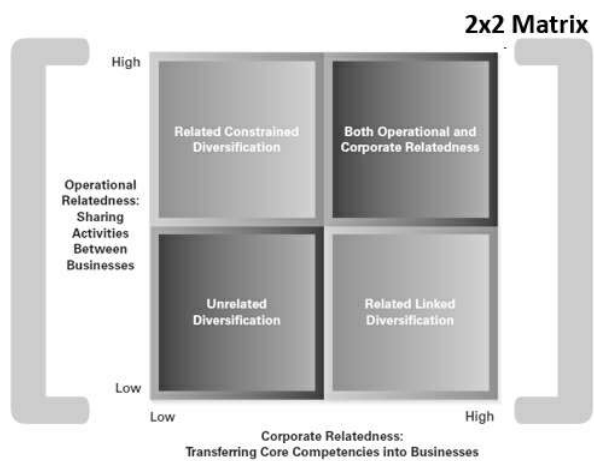

**Fig. 5.** Value-Creation Diversification Matrix adapted from [1]

**BCG Product Portfolio Matrix** (aka. *Growth-share Matrix*): This matrix which is given to the various segments within their mix of businesses [1] is formulated by the Boston Consulting Group (BCG) since 1970s [25]. It is targeted to help with long-term strategic planning, to help a business consider growth opportunities by reviewing its portfolio of products to decide where to invest, to discontinue or develop products. The decision parameters of the growth-share matrix are rate of market growth and market share and the matrix plots four strategy in a 2-by-2 matrix form as shown in Fig. 10.

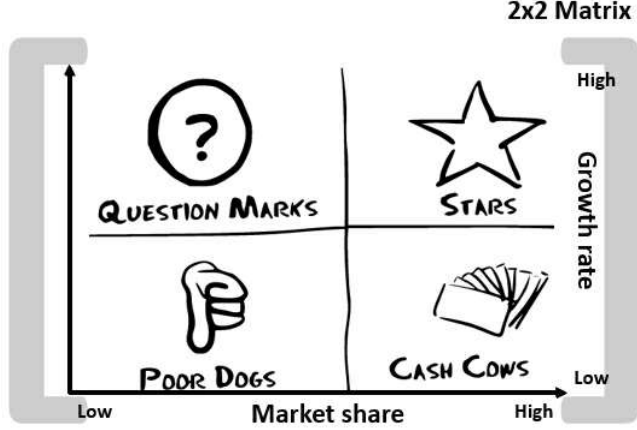

**Fig. 10.** BCG Product Portfolio Matrix [23]

It is noted a BCG growth-share matrix contains a quantitative scale of each decision parameter [10]. Therefore, the newly proposed analytical model in the paper could be easily adapted into this matrix form. Previously mentioned, actual implementation of the mathematical model into a BCG growth-share matrix is provided in Section 4.

- *Dogs* (I): The usual marketing advice here is to aim to remove any dogs from your product portfolio as they are a drain on resources.
- *Cows* (II): The simple rule here is to "Milk these products as much as possible without killing the cow!" Often mature, well-established products.
- Stars (III): The market leader though require ongoing investment to sustain. They generate more ROI than other product categories.
- *Question Marks* (IV): These products often require significant investment to push them into the star quadrant.

## 3. THE EXPLICIT MATHEMATICAL MODEL

### 3.1 Preliminaries

Let $(\Omega, \mathfrak{F}(\Omega), P)$ be probability space $\mathfrak{F}_A, \mathfrak{F}_B, \mathfrak{F}_\tau \subseteq \mathcal{F}(\Omega)$ be independent $\sigma$-subalgebras and $a_j$, $b_k$ be the level of thresholds of two strategic selection factors Suppose:

$$\mathcal{A} := \sum_{j\geq 0} a_j \varepsilon_{s_j}, \quad \mathcal{B} := \sum_{k\geq 0} b_k \varepsilon_{s_k}, s_0(=0) < s_1 < s_2 < \cdots, \text{ a.s.} \quad (3.1)$$

are $\mathfrak{F}_A$-measurable and $\mathfrak{F}_B$-measurable marked Poisson processes ($\varepsilon_a$ is a point mass at $a$) with respective intensities $\lambda_a$ and $\lambda_b$ and point independent marking. These two values are related with the market transitions. The market is observed at random times in accordance with the point process:

$$\mathcal{T} := \sum_{i\geq 0} \varepsilon_{\tau_i}, \quad \tau_0(>0)), \tau_1, \ldots, \quad (3.2)$$

which is assumed to be a delayed renewal process. A delayed renewal point process is the same as an ordinary renewal point process, except~that the first point mass is allowed to have a different one. The formula (3.2) is the formal form of a delayed renewal point process with a point mass $\varepsilon_a$ at the moment $a$ and and it becomes the sum of point masses. It is noted that point masses at $\tau_k$ are inter-arrival times between the interval $[\tau_{k-1}, \tau_k]$. Let

$$(A(t), B(t)) := \mathcal{A} \otimes \mathcal{B}([0, \tau_k]), \ k = 0, 1, \ldots, \tag{3.3}$$

be the nondecreasing random measures for strategic impacts of two decision parameters embedded over $\mathcal{T}$. With respective increments, we have:

$$(a_k, b_k) := \mathcal{A} \otimes \mathcal{B}([\tau_{k-1}, \tau_k]), \ k = 1, 2, \ldots, \tag{3.4}$$

and

$$a_0 = A_0, \ b_0 = B_0. \tag{3.5}$$

The observation process could be formalized as

$$\mathcal{A}_\tau \otimes \mathcal{B}_\tau := \sum_{k \geq 0} (a_k, b_k) \varepsilon_{\tau_k}, \tag{3.6}$$

where

$$\mathcal{A}_\tau = \sum_{i \geq 0} a_i \varepsilon_{\tau_i}, \ \mathcal{B}_\tau = \sum_{i \geq 0} b_i \varepsilon_{\tau_i}, \tag{3.7}$$

and it is with position dependent marking and with $a_k$ and $b_k$ being dependent with the notation

$$\Delta_k := \tau_k - \tau_{k-1}, \ k = 0, 1, \ldots, \ \tau_{-1} = 0, \tag{3.8}$$

and

$$\gamma(z, g, \theta, \vartheta) = \gamma_a(z, \theta) \gamma_b(g, \vartheta), \tag{3.9}$$

$$\gamma_0(z, g, \theta, \vartheta) = \gamma_a^0(z, \theta) \gamma_b^0(g, \vartheta), \tag{3.10}$$

where

$$\gamma_a(z, \theta) = \mathbb{E}\big[z^{a_j} e^{-\theta \Delta_j}\big], \ \gamma_b(g, \vartheta) = \mathbb{E}\big[g^{b_k} e^{-\vartheta \Delta_k}\big]. \tag{3.11}$$

$$\gamma_a^0(z, \theta) = \mathbb{E}\big[z^{A_0} e^{-\theta \tau_0}\big], \ \gamma_b^0(g, \vartheta) = \mathbb{E}\big[g^{B_0} e^{-\vartheta \tau_0}\big]. \tag{3.12}$$

By using the double expectation,

$$\gamma(z, g, \theta, \vartheta) = \delta(\theta + \vartheta + \lambda_a(1 - g) + \lambda_a(1 - z)), \tag{3.13}$$

and

$$\gamma_0(z, g, \theta, \vartheta) = \delta_0(\theta + \vartheta + \lambda_a(1 - g) + \lambda_b(1 - z)), \tag{3.14}$$

where

$$\delta(\theta) = \mathbb{E}\big[e^{-\theta \Delta_1}\big], \ \delta_0(\theta) = \mathbb{E}\big[e^{-\theta \tau_0}\big], \tag{3.15}$$

are the magical transform of increments $\Delta_1, \Delta_2, \ldots$. The game in this case is a stochastic process $\mathcal{A}_\tau \otimes \mathcal{B}_\tau$ describing the evolution of a stategy matrix between two strategic decision parameters A and B known to an observation process $\mathcal{T} = \{\tau_0, \tau_1, \ldots\}$. The strategic choice will be shifted when either the thresholds of the decision parameter A passes on the $j$-th observation epoch $\tau_k$ or the thresholds of the decision parameter B passes on the $k$-th observation epoch $\tau_k$. To further formalize the game, the *exit index* is introduced:

$$\mu := inf\{j : A_j = A_0 + a_1 + \cdots + a_j \geq m\}. \tag{3.16}$$

$$\nu := inf\{k : B_k = B_0 + b_1 + \cdots + b_k \geq n\}, \tag{3.17}$$

and the joint functional of the blockchain network model is as follows:

$$\begin{aligned} \Phi_{(m,n)} &= \Phi_{(m,n)}(z, g, \theta_0, \theta_1, \vartheta_0, \vartheta_1) \\ &= \mathbb{E}\left[z^\mu \cdot g^\nu \cdot e^{-\theta_0 \tau_{\mu-1}} e^{-\theta_1 \tau_\mu} \cdot e^{-\vartheta_0 \tau_{\nu-1}} e^{-\vartheta_1 \tau_\nu} \cdot \mathbf{1}_{\{A_\mu \leq m\}} \cdot \mathbf{1}_{\{B_\nu \leq n\}}\right], \end{aligned} \tag{3.18}$$

where $m$ and $n$ are the threshold of each strategic decision options. This functional will represent the status of the strategy analysis upon with the shifting time $\tau_\mu$ and $\tau_\nu$. one observation prior to this. The Theorem 1 establishes an explicit formula $\Phi_{(m,n)}$ from (3.13)-(3.14). The first exceed model by Dshahalow [14, 15] has been adopted and its operators are defined as follows:

$$\mathcal{D}_{(h,i)}[f(h,i)](x,y) := (1-x)(1-y)\sum_{h \geq 0}\sum_{i \geq 0} f(h,i)x^h y^i, \tag{3.19}$$

then

$$f(h,i) = \mathfrak{D}^{(h,i)}_{(x,y)}\left[\mathcal{D}_{(x,y)}\{f(h,i)\}\right], \tag{3.20}$$

where $\{f(h,i)\}$ is a sequence, with the inverse

$$\mathfrak{D}^{(h,i)}_{(x,y)}(\bullet) = \begin{cases} \left(\frac{1}{h! \cdot i!}\right) \lim_{(x,y) \to 0} \frac{\partial^h \partial^i}{\partial x^h \partial y^i} \frac{1}{(1-x)(1-y)}(\bullet), & h \geq 0, \ i \geq 0, \\ 0, & \text{otherwise.} \end{cases} \tag{3.21}$$

The functional $\mathfrak{D}$ is defined on the space of all analytic functions at 0 and it has the following properties:

(i) $\mathfrak{D}^j_x$ is a linear functional with fixed points at constant functions,

(ii) $\mathfrak{D}^m_x \sum_{j=0}^{\infty} a_j x^j = \sum_{j=0}^{m} a_j.$ \hfill (3.22)

**Theorem 1:** the functional $\Phi_{(m,n)}$ of the process of (3.18) satisfies following expression:

$$\Phi_{(m,n)} = \mathfrak{D}^{(m,n)}_{(x,y)}\left[\Psi(x,y)\right]. \tag{3.23}$$

where

$$\Psi(x,y) = \frac{(\delta_{\theta_1} - \varphi_a)(\delta_{\vartheta_1} - \varphi_b)(z - z\Gamma^0_a \Gamma_a + \Gamma^0_a)(g - g\Gamma^0_b \Gamma_b + \Gamma^0_b)}{(1 - z\Gamma_a)(1 - g\Gamma_b)} \tag{3.24}$$

and

$$\Gamma_a = \gamma_a(x, \theta_0 + \theta_1) = \delta_{\theta_0} \cdot \varphi_a, \tag{3.25}$$
$$\varphi_a = \gamma_a(x, \theta_1), \tag{3.26}$$

$$\delta_\theta = \delta(\theta),\ \delta_\theta^0 = \delta_0(\theta), \tag{3.27}$$
$$\Gamma_b = \gamma_b(y, \vartheta_0 + \vartheta_1) = \delta_{\vartheta_0} \cdot \varphi_b, \tag{3.28}$$
$$\varphi_b = \gamma_b(y, \vartheta_1). \tag{3.29}$$

From (3.138) and (3.23)-(3.24), we can find the PGFs (probability generating functions) of the *exit index* $\mu$ (and $\nu$) :

$$\mathbb{E}[z^\mu] = \Phi_{(m,n)}(z, 1, 0, 0, 0, 0), \tag{3.30}$$
$$\mathbb{E}[g^\mu] = \Phi_{(m,n)}(1, g, 0, 0, 0, 0), \tag{3.31}$$

$$\mathbb{E}\left[e^{-\theta_0 \tau_{\mu-1}}\right] = \Phi_{(m,n)}(1, 1, \theta_0, 0, 0, 0), \tag{3.32}$$
$$\mathbb{E}\left[e^{-\theta_1 \tau_\mu}\right] = \Phi_{(m,n)}(1, 1, 0, \theta_1, 0, 0), \tag{3.33}$$

$$\mathbb{E}\left[e^{-\vartheta_0 \tau_{\nu-1}}\right] = \Phi_{(m,n)}(1, 1, \vartheta_0, 0, 0, 0), \tag{3.34}$$
$$\mathbb{E}\left[e^{-\vartheta_1 \tau_\nu}\right] = \Phi_{(m,n)}(1, 1, 0, \vartheta_1, 0, 0). \tag{3.35}$$

### 3.2. General Matrix Frameworks for Strategy Formulation

Recalling from the previous section, the matrix form is atypical way to describe the set of strategic choices based on the decision parameters (i.e., the category of the strategy input) in the strategy formulation. Most strategy formulation could be described by the matrix. The section deals the conventional 2-by-2 matrix form to combine the fluctuation theory for determine the proper strategic decision making. The stakeholder mapping \cite{02} in Fig. 3, the product-market matrix in Fig. 10, the business level strategy in Fig. 6 and the BCG product portfolio in Fig. 9 are intuitively described by a 2-by-2 matrix form.

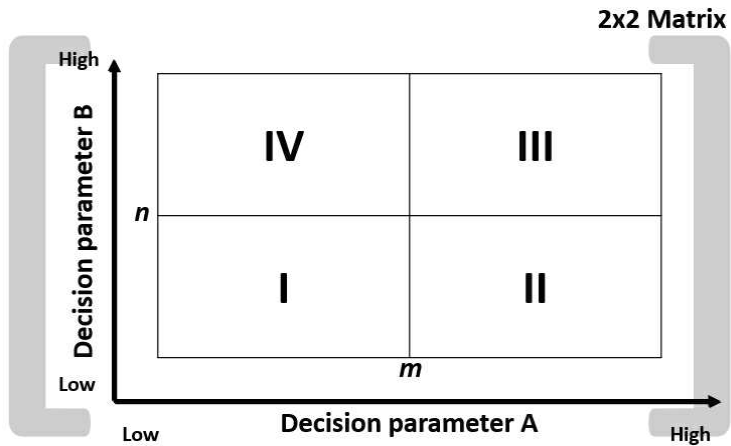

**Fig. 11.** 2x2 Conventional Strategy Matrix

Generally, the 2-*by*-2 matrix provides the four strategic choices (I-IV) and the optimal strategic choice depends on a present or a future position of the company. Let us assume that the decision parameters are quantitative and the thresholds are exists which are equivalent with $m$ and $n$ from (3.16)-(3.17). The strategy I becomes be the best when $A_j \leq m$, $B_k \leq n$ and strategy II is the best when $A_j > m$, $B_k \leq n$. Formally speaking, the best strategic choice could be chosen as follows:

$$BSF(A_j, B_k) := \begin{cases} \text{I}, & A_j \leq m,\ B_k \leq n, \\ \text{II}, & A_j > m,\ B_k \leq n, \\ \text{III}, & A_j > m,\ B_k > n, \\ \text{IV}, & A_j \leq m,\ B_k > n, \end{cases} \quad (3.36)$$

from (3.16)-(3.17). The moment of selecting the strategic choice becomes the critical matter and the first exceed model could analytically solve when the strategic decision maker takes the action. Generally, the moment of shifting the strategy is the time when the decision parameters pass their thresholds $\{m, n\}$ which could be determined analytically from (3.16) and (3.17). But a strategic choice could be shifted before passing the strategic thresholds which are $\tau_{\mu-1}$ and $\tau_{\nu-1}$ instead of the first exceed moments $\tau_\mu$ and $\tau_\nu$.

### 3.3. Find The Moment of The Strategy Shifts

It is assumed that the observation process has the memoryless properties which might be a special condition but very practical for actual implementation on a decision making system. It implies that the observation process dose not have any dependence with the strategic decision parameters. Recall from (3.19), the operator is defined as follows:

$$F(x) = \mathcal{D}_k\left[g(k)\right](x) := (1-x)\sum_{k \geq 0} g(k) x^k, \quad (3.37)$$

$$\mathcal{D}_{(j,k)}\left[f_1(j)f_2(k)\right](x, y) := (1-x)(1-y)\sum_{j \geq 0}\sum_{k \geq 0} g_1(j)g_2(k) x^j y^k$$
$$= \mathcal{D}_j\left[g_1(j)\right]\mathcal{D}_k\left[g_2(k)\right],$$

then

$$g(j, k) = \mathfrak{D}_{(x,y)}^{(j,k)}\left[\mathcal{D}_{(j,k)}\{g(j, k)\}\right], \quad (3.38)$$

$$g_1(j)g_2(k) = \mathfrak{D}_x^j[\mathcal{D}_j\{g_1(j)\}]\mathfrak{D}_y^k[\mathcal{D}_k\{g_2(k)\}], \quad (3.39)$$

where $\{f(x), (f_1(x)f_2(y))\}$ are a sequence, with the inverse (3.21) and

$$\mathfrak{D}_x^k(\bullet) = \begin{cases} \frac{1}{k!} \lim_{x \to 0} \frac{\partial^k}{\partial x^k} \frac{1}{(1-x)}(\bullet), & m \geq 0, \\ 0, & \text{otherwise,} \end{cases} \quad (3.40)$$

and

$$\mathfrak{D}_{(x,y)}^{(j,k)}[F_1(u)F_2(v)] = \mathfrak{D}_x^j[F_1(x)]\mathfrak{D}_y^k[F_2(y)]. \quad (3.41)$$

The marginal mean of $\tau_\mu$ and $\tau_\nu$ are the moment of the strategy chances and it could be straight forward once the exit index is found. Each *exit index* of the decision parameters A and B could be found as the **Lemma** 1 and 2:

**Lemma 1:** The probability generating function (PGF) of the *exit index* for the decision parameter $A$ could found as follows:

$$\mathbb{E}[z^\mu] = z + \left(\frac{1-\kappa}{1+\widetilde{\delta}_0 \cdot \lambda_a}\right)\left[\sum_{j=0}^{m}\left(\frac{\widetilde{\delta}_0 \cdot \lambda_a}{1+\widetilde{\delta}_0 \cdot \lambda_a}\right)^j\right] + \left(\frac{\kappa}{1+\widetilde{\delta}_0 \cdot \lambda_a}\right)\left[\sum_{j=0}^{m}\left(\frac{\widetilde{\delta} \cdot \lambda_a}{1+\widetilde{\delta} \cdot \lambda_a}\right)^j\right]$$

$$-\frac{(1-z)z}{\left(1+\widetilde{\delta}\cdot\lambda_a\right)-z}\left[\sum_{k=0}^{m}\left(\frac{\left(\widetilde{\delta}\cdot\lambda_a\right)}{\left(1+\widetilde{\delta}\cdot\lambda_a\right)-z}\right)^{k}\right], \tag{3.42}$$

and the PGF of the exit index for decision parameter B could be found as the
**Lemma 2:**

$$\mathbb{E}[g^{\nu}] = g + \left(\frac{1-\kappa^{1}}{1+\widetilde{\delta_0}\cdot\lambda_b}\right)\left[\sum_{j=0}^{n}\left(\frac{\widetilde{\delta_0}\cdot\lambda_b}{1+\widetilde{\delta_0}\cdot\lambda_b}\right)^{j}\right] + \left(\frac{\kappa^{1}}{1+\widetilde{\delta}\cdot\lambda_b}\right)\left[\sum_{j=0}^{n}\left(\frac{\widetilde{\delta}\cdot\lambda_b}{1+\widetilde{\delta}\cdot\lambda_b}\right)^{j}\right]$$

$$-\frac{(1-g)g}{\left(1+\widetilde{\delta}\cdot\lambda_b\right)-g}\left[\sum_{k=0}^{n}\left(\frac{\left(\widetilde{\delta}\cdot\lambda_b\right)}{\left(1+\widetilde{\delta}\cdot\lambda_b\right)-g}\right)^{k}\right] \tag{3.43}$$

where

$$\Gamma_b^0 = \gamma_b^0(y, \vartheta_0 + \vartheta_1) = \frac{1}{\left(1+\widetilde{\delta_0}\cdot(\lambda_a+\vartheta_0+\vartheta_1)\right)-\widetilde{\delta_0}\cdot\lambda_b y} = \frac{\beta_b^0}{1-\alpha_b^0\cdot y}, \tag{3.44}$$

$$\Gamma_b = \gamma_b(x, \vartheta_0 + \vartheta_1) = \frac{1}{\left(1+\widetilde{\delta}\cdot(\lambda_b+\vartheta_0+\vartheta_1)\right)-\widetilde{\delta}\cdot\lambda_b y} = \frac{\beta_b}{1-\alpha_b\cdot y}, \tag{3.45}$$

$$\alpha_b = \frac{\widetilde{\delta}\cdot\lambda_b}{1+\widetilde{\delta}\cdot\lambda_b},\ \beta_b = \frac{1}{1+\widetilde{\delta}\cdot\lambda_b},\ \alpha_b^0 = \frac{\widetilde{\delta_0}\cdot\lambda_b}{1+\widetilde{\delta_0}\cdot\lambda_b},\ \beta_b = \frac{1}{1+\widetilde{\delta_0}\cdot\lambda_b}, \tag{3.46}$$

$$\alpha_b = \frac{\widetilde{\delta}\cdot\lambda_b}{1+\widetilde{\delta}\cdot\lambda_b},\ \beta_b = \frac{1}{1+\widetilde{\delta}\cdot\lambda_b},\ \alpha_b^0 = \frac{\widetilde{\delta_0}\cdot\lambda_b}{1+\widetilde{\delta_0}\cdot\lambda_b},\ \beta_a = \frac{1}{1+\widetilde{\delta_0}\cdot\lambda_b}, \tag{3.48}$$

$$\kappa^{1} = \frac{1}{\lambda_b\left(\widetilde{\delta_0}-\widetilde{\delta}\right)},\ \widetilde{\delta_0} = \mathbb{E}[\tau_0],\ \widetilde{\delta} = \mathbb{E}[\Delta_1]. \tag{3.50}$$

From (3.42),

$$\mathbb{E}[\mu] = \frac{1}{\widetilde{\delta}\cdot\lambda_a},\ \mathbb{E}[\nu] = \frac{1}{\widetilde{\delta}\cdot\lambda_b}, \tag{3.52}$$

and

$$\mathbb{E}[\tau_\mu] = \widetilde{\delta_0} + \widetilde{\delta}\cdot\mathbb{E}[\mu-1] = \widetilde{\delta_0} + \frac{1}{\lambda_a} - \widetilde{\delta}, \tag{3.53}$$

$$\mathbb{E}[\tau_\nu] = \widetilde{\delta_0} + \widetilde{\delta}\cdot\mathbb{E}[\nu-1] = \widetilde{\delta_0} + \frac{1}{\lambda_b} - \widetilde{\delta}. \tag{3.54}$$

### 3.4. Useful Tips for Theoretical Modeling

Since we are dealing with the mathematical approach, it is impossible to apply a theoretical model into a practical implementation without adjustments. Although all strategy formulations in the literature review are described as a matrix form, not all of them have quantitative decision parameters. It is noted that all decision parameters should be quantitative and the thresholds should be measurable to be applied into an analytical model. Additionally, the scale of decision parameters might be modified for sustaining the linearity of the decision parameter values. In this newly proposed mathematical strategy formulation model, it should be assumed that the process for each decision parameter is the Poisson compound process and this is the most mandatory condition that makes the mathematical model analytically solvable. Instead of the Poisson process, a generally distributed random process of decision parameters might be considered when the related data are possibly obtained. Similarly, a numerical approach could be considered instead of an analytical approach if it is feasible to collecting suitable real data in a certain way. Lastly, the observation process might have the memoryless properties which implies that the observation process does not contain any past information.

## 4. BCG GROWTH-SHARE MATRIX

In the literature review, the BCG growth-share matrix contains the quantitative scale of each decision parameter. Using the BCG growth-share matrix with the quantitative scale has been shown in Fig. 12. It is one of intuitive way to portray a company's portfolio strategy [11]. According to Hedley [11], the separating areas of high and low relative competitive position is set at 1.5 times. If a product has relative strengths of this magnitude to ensure that it will have the dominant position needed to be Stars or Cows. In the BGC growth-share matrix model, $A_j$ represents the transformed random value from the a market share $c_j$ of a product (or a company) at $j$-th epoch and $B_k$ represents a growth rate (%) of a market at $k$-th epoch. Both are assumed to be monotone nondecreasing until a strategy is shifted. The scale of the BCG matrix should be modified to adapt the mathematical model and the scale of the relative competitive position is mapped as follows:

$$A_j(c_j) = \left\lfloor 10^2 log\left\{10 \cdot \left(\frac{c_j}{1-c_j}\right)\right\}\right\rfloor, c_j \in [0,1), \tag{4.1}$$

and the market share $c_j$ is transformed into the relative competitive ratio with the revised scale (i.e., the x-axis in Fig. 12) and the modified mapping scale could be shown in Fig. 13.

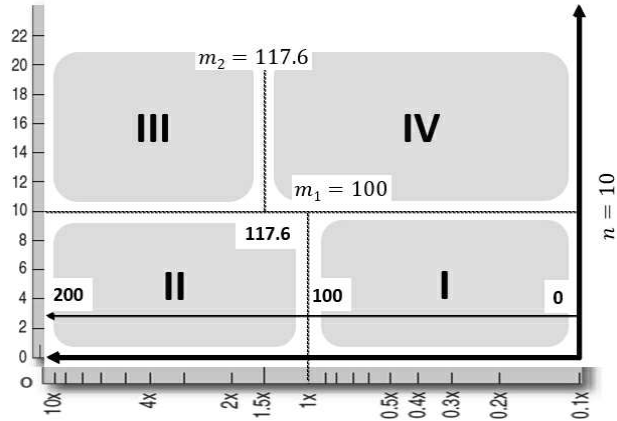

**Fig. 13.** Scale Modification of the BCG matrix

From (3.36), the best strategy for the BCG matrix is determined as follows:

$$BSF(A_j, B_k) := \begin{cases} \textit{Dogs (I)}, & A_j \leq 100, B_k \leq 10, \\ \textit{Cows (II)}, & A_j > 100, B_k \leq 10, \\ \textit{Stars (III)}, & A_j > 118, B_k > 10, \\ \textit{Question Marks (IV)}, & A_j \leq 117, B_k > 10, \end{cases} \tag{4.1}$$

and the optimal moments to shift the strategy $\mathbb{E}[\tau_\mu]$ and $\mathbb{E}[\tau_\nu]$ are determined from (3.52) and (3.53).

## 5. CONCLUSION

The main contribution of this paper is establishing the theoretical framework of the strategy formulations by providing the explicit forms from a conventional strategy formulation matrix. The mathematical analysis which includes the explicit functionals for finding the optimal moments of the strategy shifts and strategy executions are fully deployed in this paper. This analytical approach supports the theoretical background of the strategic choice to make the best decisions from advanced mathematical strategy formulations. Additionally, this newly proposed analytical model has been applied into the BCG growth-share matrix to demonstrate the actual implementation of the model. This research shall be helpful for whom is looking for scientific strategy executions in real business matters.